# FCC-ee accelerator parameters, performance and limitations

Mike Koratzinos[a] on behalf of the FCC-ee study

[a] *University of Geneva, Switzerland*

**Abstract**

CERN has recently launched the Future Circular Collider (FCC) study to deal with all aspects of an ambitious post-LHC possible programme. The emphasis of the study is on a 100 TeV proton collider to be housed in a 80-100 km new ring in the Geneva region. An electron machine will also be considered as a possible intermediate first step (FCC-ee). The study benefits from earlier work done in the context of TLEP and has already published a parameter table, to serve as the basis for the work to be done. The study aims to publish a conceptual design report at around 2018. The recent discovery of a light Higgs boson has opened up considerable interest in circular $e^+e^-$ Higgs factories around the world. FCC-ee is capable of very high luminosities in a wide centre-of-mass ($E_{CM}$) spectrum from 90 to 350 GeV. This allows the very precise study of the Z, W and H bosons as well as the top quark, allowing for meaningful precision tests of the closure of the Standard Model.

## 1. Introduction

The discovery of Higgs boson, one of the last missing pieces of the standard model, opens the possibility of stringent closure tests of the theory. The precision, clean environment and high luminosity of $e^+e^-$ colliders make them perfect vehicles for such tests. In particular, circular $e^+e^-$ machines can deliver high luminosities to probe all interesting particles around the electroweak scale: the Z, the W and the Higgs bosons, as well as the top quark.

The FCC-ee is a study for a circular $e^+e^-$ collider that can operate at beam energies from 45GeV to 175GeV with unprecedented luminosity. The size of the ring will be 80 - 100 kms (current baseline) and it is to be situated in the Geneva region to profit from CERN's infrastructure. Four experiments are envisaged. Synchrotron radiation power dissipation is fixed at 50MW per beam. Separate beam pipes for $e^+$ and $e^-$ are envisaged, giving considerable flexibility to the design, as well as allowing a large number of bunches. The accelerator will use a top-up ring (referred to as the booster hereafter) to keep the luminosity close to its maximum value. The design study team has published a first parameter set [1] to be used as a basis for the study, which is summarised in Table 1.

## 2. Maximum luminosity of a circular collider

The luminosity of a circular collider is given by

$$\mathcal{L} = \frac{3}{8\pi}\frac{e^4}{r_e^4}P_{tot}\frac{\rho}{E_0^3}\xi_y\frac{R_{hg}}{\beta_y^*}$$

where $r_e$ and $e$ are the classical radius of the electron and its charge, $P_{tot}$ the total SR power dissipated by one beam, $\rho$ the bending radius, $E_0$ the beam energy, $\xi_y$ vertical beam-beam parameter, $\beta_y^*$ the vertical beta function at the interaction point and $R_{hg}$ the geometric hourglass factor.

The maximum achievable $\xi_y$ depends on if a specific machine is beam-beam or beamstrahlung dominated [2]. The beam-beam limit depends on the damping decrement, the amount of energy loss when electrons move from one IP to the next. The LEP data has been used to derive this number at FCC-ee energies and machine circumference following the formulation in [3].

Table 1: FCC-ee baseline parameters compared to LEP1 and LEP2

| | (LEP1) | (LEP2) | Z | W | H | tt |
|---|---|---|---|---|---|---|
| Circumference [km] | 26.7 | 26.7 | 100 | 100 | 100 | 100 |
| Bending radius [km] | 3.1 | 3.1 | 11 | 11 | 11 | 11 |
| Beam energy [GeV] | 45.4 | 104 | 45.5 | 80 | 120 | 175 |
| Beam current [mA] | 2.6 | 3.04 | 1450 | 152 | 30 | 6.6 |
| Bunches / beam | 12 | 4 | 16700 | 4490 | 1360 | 98 |
| Bunch population [$10^{11}$] | 1.8 | 4.2 | 1.8 | 0.7 | 0.46 | 1.4 |
| Transverse emittance $\varepsilon$ | | | | | | |
| - Horizontal [nm] | 20 | 22 | 29.2 | 3.3 | 0.94 | 2 |
| - Vertical [pm] | 400 | 250 | 60 | 7 | 1.9 | 2 |
| Momentum comp. [$10^{-5}$] | 18.6 | 14 | 18 | 2 | 0.5 | 0.5 |
| Betatron function at IP $\beta^*$ | | | | | | |
| - Horizontal [m] | 2 | 1.2 | 0.5 | 0.5 | 0.5 | 1 |
| - Vertical [mm] | 50 | 50 | 1 | 1 | 1 | 1 |
| Beam size at IP $\sigma^*$ [µm] | | | | | | |
| - Horizontal | 224 | 182 | 121 | 41 | 22 | 45 |
| - Vertical | 4.5 | 3.2 | 0.25 | 0.084 | 0.044 | 0.045 |
| Bunch length [mm] | | | | | | |
| - Synchrotron radiation | 8.6 | 11.5 | 1.64 | 1.01 | 0.81 | 1.16 |
| - Total | 8.6 | 11.5 | 2.56 | 1.49 | 1.17 | 1.49 |
| Energy loss / turn [GeV] | 0.12[1] | 3.34 | 0.03 | 0.33 | 1.67 | 7.55 |
| SR power / beam [MW] | 0.3[1] | 11 | 50 | 50 | 50 | 50 |
| Total RF voltage [GV] | 0.24 | 3.5 | 2.5 | 4 | 5.5 | 11 |
| RF frequency [MHz] | 352 | 352 | 800 | 800 | 800 | 800 |
| Longitudinal damping time $\tau_E$ [turns] | 371 | 31 | 1320 | 243 | 72 | 23 |
| Energy acceptance RF [%] | 1.7 | 0.8 | 2.7 | 7.2 | 11.2 | 7.1 |
| Synchrotron tune Qs | 0.065 | 0.083 | 0.65 | 0.21 | 0.096 | 0.10 |
| Polarization time $\tau_p$ [min] | 252 | 4 | 11200 | 672 | 89 | 13 |
| Hourglass factor H | 1 | 1 | 0.64 | 0.77 | 0.83 | 0.78 |
| Luminosity/IP [$10^{34}$ cm$^{-2}$s$^{-1}$] | 0.002 | 0.012 | 28.0 | 12.0 | 6.0 | 1.8 |
| Beam-beam parameter | | | | | | |
| - Horizontal | 0.044 | 0.040 | 0.031 | 0.060 | 0.093 | 0.092 |
| - Vertical | 0.044 | 0.040 | 0.030 | 0.059 | 0.093 | 0.092 |
| Luminosity lifetime [min] | 1750 | 434 | 298 | 73 | 29 | 21 |

[1] Does not take into account the contribution of damping and emittance wigglers

The beamstrahlung limit [4] is due to the fact that at high energies and luminosities beamstrahlung, the synchrotron radiation emitted by an incoming electron in the collective electromagnetic field of the opposite bunch at an interaction point, reduces beam lifetimes to values where the top-up injector cannot cope. The effect of beamstrahlung is very implementation specific and can be mitigated by small vertical emittance and large momentum acceptance.

The two regimes, for the specific implementation of FCC in [1], can be seen in Figure 1. Such a machine would be beamstrahlung dominated at 175GeV, but at lower energies will be beam-beam dominated.

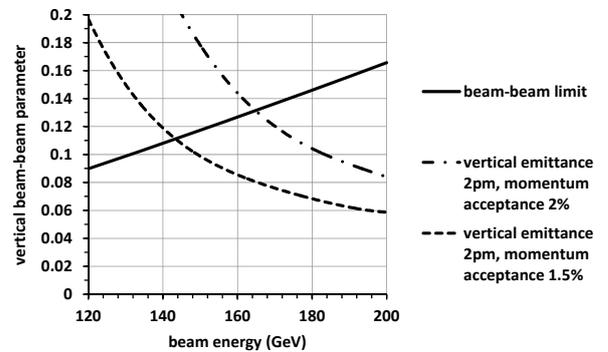

Figure 1: The beam-beam and beamstrahlung limits for a collider with parameters as in [1]. The beamstrahlung curves have been obtained assuming a beam lifetime of 300 seconds and two different momentum acceptance values of 1.5% and 2%.

## 3. The FCC-ee parameter set

The vertical beta function at the IP is limited by the final focus layout and chromatic correction. A very small vertical β* of 1 mm is set as baseline parameter. It however remains a factor of three times larger than the target value of SuperKEKB.

The arc lattice design uses a FODO cell with a length of 50 m for operation at high energies (120 GeV and above). This gives the desired low horizontal emittance at these energies. At lower energies, such low emittance increases the number of bunches without any gain in luminosity, so the phase advance is lowered by increasing the length of the cell in units of the basis cell. For the W working point the cell length is doubled, for the Z working point it is increased by a factor of 6. The ratio of horizontal to vertical emittance is assumed to be 500 (1000 for $t\bar{t}$ running), which is aggressive but should be possible with modern beam instrumentation.

The design of a final focus system with an energy acceptance of around 2% is very challenging and work has already started.

An 800 MHz RF system is considered as the baseline option to maintain sufficiently short bunch lengths. The total RF voltage at the $t\bar{t}$ threshold will be in the range of 11 to 12 GV. If a gradient of 20 MV/m is assumed, the length of the RF system will be 600 m for each beam. The option of using a lower frequency RF system will be part of the study.

The short lifetimes from radiative Bhabha scattering and from beamstrahlung require continuous top-up injection. The FCC-ee collider will therefore be operated at constant energy and must be continuously filled by a cycling booster ring. The booster is installed in the FCC tunnel and has its own RF system, with same length and total voltage as the main ring. The currents in the booster will be at the level of a few percent of the full current of the main ring, therefore the required RF power is correspondingly lower. The booster will be cycling between its injection energy of 10 - 40 GeV and the operating energy of FCC-ee. The maximum repetition rate is currently estimated to be around 0.1 Hz. The required flux of $e^+$ and of $e^-$ is estimated to be $2\times10^{12}$ particles per second for each species.

Beam polarization is an important parameter for Z and W operation. Transverse polarization is mandatory for achieving precise energy calibration, an important part of the physics programme at the W and Z. The measurement is done using the resonant depolarization technique, which provides an exceptionally accurate measurement of the beam energy, to the level of 0.1 MeV or better [5]. To achieve such accuracy, continuous monitoring of the beam energy with polarized non-colliding bunches is called for. With the very small momentum compaction factor of FCC-ee the range of energy variations is expected to be larger than 100 MeV (tides, geological movements etc). The extrapolation of the average beam energy (given by the resonant depolarization method) to the centre of mass energy at the IPs requires an excellent understanding of all sources contributing to local energy changes (such as RF voltages and phases, etc.).

The current baseline assumes head-on or nearly head on collisions with small crossing angles (0.1 - 1 mrad). A crab-waist scheme has recently been studied [4] and seems to offer substantial gains in luminosity for Z and W running, as can be seen in Table 2. Such a scheme will be studied in more detail, and it may become part of a future baseline.

Table 2: Crab-waist scheme proposal and expected gain in luminosity. Values shown are derived from a quasi-strong-strong simulation.

| ECM Energy [GeV] | 45.5 | | 80 | |
|---|---|---|---|---|
| Collision scheme | Head-on | Crab waist | Head-on | Crab waist |
| $N_p$ [$10^{11}$] | 1.8 | 1.0 | 0.7 | 4.0 |
| θ [mrad] | 0 | 30 | 0 | 30 |
| $σ_z$ (SR/total) [mm] | 1.6/3.0 | 2.8/7.6 | 1.0/1.8 | 4.1/11.6 |
| $ε_x$ [nm] | 29.2 | 0.14 | 3.3 | 0.44 |
| $ε_y$ [pm] | 60 | 1.0 | 7.0 | 1.0 |
| $ξx/ξy$ | 0.03/0.03 | 0.02/0.14 | 0.06/0.06 | 0.02/0.20 |
| L [$10^{34}$ cm$^{-1}$s$^{-1}$] | 17 | 180 | 13 | 45 |